\documentclass[12pt]{article}

\pdfoutput=1
\usepackage{color}
\usepackage{epsfig, palatino}
\usepackage{mathrsfs}
\usepackage{ae} 
\usepackage[T1]{fontenc}
\usepackage[ansinew]{inputenc}
\usepackage{amsmath}
\usepackage{amssymb}
\usepackage{graphicx}
\usepackage{color}
\definecolor{darkblue}{cmyk}{0.9,0.9,0,0}
\usepackage[colorlinks=true,linkcolor=darkblue,citecolor=darkblue,urlcolor=darkblue]{hyperref}
\usepackage{hyperref}
\usepackage{varioref}
\usepackage{makeidx}
\usepackage[english]{babel}
\usepackage{array}
\usepackage{multirow}

\newcommand{\comment}[1]{}

\newcommand{\beq}{\begin{equation}}
\newcommand{\eeq}{\end{equation}}
\newcommand{\beqq}{\begin{equation*}}
\newcommand{\eeqq}{\end{equation*}}
\newcommand\beqa{\begin{eqnarray}}
\newcommand\eeqa{\end{eqnarray}}
\newcommand\beqaa{\begin{eqnarray*}}
\newcommand\eeqaa{\end{eqnarray*}}
\newcommand\bea{\begin{array}}
\newcommand\eea{\end{array}}

\newcommand{\neqa}{\nonumber\end{eqnarray}}

\renewcommand{\d}{\partial}

\newcommand{\<}{{\langle}}
\renewcommand{\>}{{\rangle}}

\newcommand{\re}{\relax{\rm I\kern-.18em R}}

\renewcommand{\sp}{p\hspace{-.40em}/}

\definecolor{darkgreen}{rgb}{0.0, 0.45, 0.0}

\def\XXint#1#2#3{{\setbox0=\hbox{$#1{#2#3}{\int}$}
\vcenter{\hbox{$#2#3$}}\kern-.5\wd0}}

\def\su2{{SU(2)}}

\def\[{\left[}
\def\]{\right]}

\def\({\left(}
\def\){\right)}
\def\[{\left[}
\def\]{\right]}

\def\<{\langle}
\def\>{\rangle}

\def\i2{\frac{i}{2}}

\def\spi{\relax{\rm \pi\kern-0.5em /}}
\def\sA{\relax{\rm A\kern-0.5em /}}
\def\sp{\relax{\rm p\kern-0.5em /}}
\def\sd{\relax{\rm \d\kern-0.5em /}}
\def\sk{\relax{\rm k\kern-0.5em /}}
\def\sn{\relax{\rm n\kern-0.5em /}}
\def\sl{\relax{\rm l\kern-0.5em /}}
\def\sP{\relax{\rm P\kern-0.7em /}}
\def\sBethe{\relax{\rm \Bethe\kern-0.5em /}}

\def\2F1{\,_2{\rm F}_1}

\makeindex

\begin{document}

\thispagestyle{empty}

\renewcommand{\thefootnote}{\fnsymbol{footnote}}
\setcounter{page}{1}
\setcounter{footnote}{0}
\setcounter{figure}{0}


\begin{center}
$$$$
{\Large\textbf{\mathversion{bold}
Extracting OPE coefficient of Konishi at four loops
}\par}

\vspace{1.0cm}

\textrm{Vasco Goncalves}
\\ \vspace{1.2cm}
\footnotesize{\textit{
ICTP South American Institute for Fundamental Research Instituto de Fisica Teorica,
UNESP - Univ. Estadual Paulista Rua Dr. Bento T. Ferraz 271, 01140-070, Sao Paulo,
SP, Brasil
}  
\vspace{4mm}
}

\par\vspace{1.5cm}

\textbf{Abstract}\vspace{2mm}
\end{center}
We compute in this short note the OPE coefficient of two $20'$ operators and the Konishi. We use the OPE decomposition of a four point function of four $20'$ operators and the method of asymptotic expansions to compute the integrals at the order that it is needed.

\noindent

\setcounter{page}{1}
\renewcommand{\thefootnote}{\arabic{footnote}}
\setcounter{footnote}{0}

\setcounter{tocdepth}{2}

 \def\nref#1{{(\ref{#1})}}

\newpage
\newpage

\tableofcontents

\parskip 5pt plus 1pt   \jot = 1.5ex

\section{Introduction}
Correlation functions of local operators in a CFT are completely determined by dimensions of operators and their OPE coefficients. Over the last years there has been a significant progress in computing the dimensions and OPE coefficients of local operators in $\mathcal{N}=4$ SYM. The integrability of the planar sector of this particular CFT  has allowed the determination of the spectrum of single trace operators at any value of the coupling. Recently, it was proposed a method (hereafter called the hexagon approach) to compute OPE coefficients of single trace operators  at any value of the coupling\cite{C123Paper}. This new approach has passed several non-trivial checks \cite{C123Paper,Basso:2015eqa,Eden:2015ija}. At weak coupling, there are new features appearing at each loop order and in the past it was useful to have these OPE coefficients computed by other means in order to check the correctness of the integrability result. The interest in the four loop stems from the appearence of a new effect in the hexagon approach due to wrapping effects \cite{C123Paper,Basso:2015eqa}. Thus, reproducing the result of this note will be an important non-trivial check of the integrability computation.

In this note we compute the OPE coefficient of two $20'$ operators and the Konishi operator at four loop level by doing the OPE decomposition of a four point function  $20'$ operators. This four point function is known only at the integrand level, so to do the OPE we will use the method of asymptotic expansions to compute the integrand in OPE limit.  This method has already been implemented in the past to determine the OPE coefficient at three loops \cite{Eden:2012rr}.  

In the next section we will define the four point function that we will be working with. Then  we briefly review the method of asymptotic expansions and finally we extract the OPE coefficient by considering a limit of the four point function.

\section{Four point function and OPE limit}
In $\mathcal{N}=4$ SYM there are special operators (often called protected) that do not receive corrections to their dimension and OPE coefficients or in other words, their two and three point function are the same at any coupling. However, a four point function of these operators does get corrected. One way to understand this is by writing the four point function as a sum of two three point functions
\begin{align}
\langle \mathcal{O}(x_1) \mathcal{O}(x_2) \mathcal{O}(x_3) \mathcal{O}(x_4) \rangle =\frac{1}{(x_{12}^2x_{34}^2)^{\Delta_{\mathcal{O}}}} \sum_{k}c_{\mathcal{O}\mathcal{O}\mathcal{O}_k}^2G_{\Delta_k,J_k}(u,v)
\end{align}
where $c_{\mathcal{O}\mathcal{O}\mathcal{O}_k}$ is an OPE coefficient, $G_{\Delta,J}(u,v)$ is a conformal block (that resums the contribution of a conformal family to a four point function) and $u$ and $v$ are cross ratios. In general the OPE coefficient $c_{\mathcal{O}\mathcal{O}\mathcal{O}_k}$ depends on the coupling and consequently the four point function also depends on it. 

Of all protected operators there is one, denoted by $\mathcal{O}_{20'}$, that is built out of scalars and sits in the same supermultiplet as the Lagrangian 
\begin{align}
 \mathcal{O}(x,y) = Y_{I}Y_{J}\mathcal{O}_{20'}^{IJ}(x)=Y_{I}Y_{J}\textrm{tr}\left(\Phi^{I}(x)\Phi^{J}(x)\right)\,, \ \ \ \ \ Y^2= Y_{I}Y_{I}=0. 
\end{align}
where the null variables $Y$ insure that the operator is symmetric and traceless in the $R$-charge indices. The structure of the four point function of $20'$ operators can be written in a compact form\cite{Eden:2000bk}
\begin{align}
&G_4=\langle \mathcal{O}(x_1,y_1)\dots \mathcal{O}(x_4,y_4) \rangle =  \sum_{l=0}^{\infty}a^{l}G_{4}^{(l)}(1,2,3,4),
\end{align}
with the tree level result given by
\begin{align}
&G^{(0)}(1,2,3,4)  = \ \frac{(N^2-1)^2}{4 (4 \pi^2)^4}\biggl(\frac{y_{12}^4 y_{34}^4}{x_{12}^4 x_{34}^4} + \frac{y_{13}^4 y_{24}^4}{x_{13}^4 x_{24}^4}  +\frac{y_{14}^4 y_{23}^4}{x_{14}^4 x_{23}^4}  \biggr) \notag \\
&\ +\frac{N^2-1}{(4 \pi^2)^4} \biggl(\frac{y_{12}^2 y_{23}^2 y_{34}^2 y_{41}^2}{x_{12}^2 x_{23}^2 x_{34}^2 x_{41}^2} + \frac{y_{12}^2 y_{24}^2 y_{43}^2 y_{31}^2}{x_{12}^2 x_{24}^2 x_{43}^2 x_{31}^2}  + \frac{y_{13}^2 y_{32}^2 y_{24}^2 y_{41}^2}{x_{13}^2 x_{32}^2 x_{24}^2 x_{41}^2}  \biggr)\,, y_{ij}=Y_i\cdot Y_j
\end{align}
and the loop level by
\begin{align}
G_4^{(l)}= \frac{2(N_c^2-1)}{(4\pi^2)^4}&R\frac{x_{12}^2x_{13}^2x_{14}^2x_{23}^2x_{24}^2x_{34}^2}{l!(-4\pi^2)^l}\int d^4x_5\dots d^4x_{4+l}f^{(l)}(x_1,\dots,x_{4+l}),\, \ \ (\textrm{for }\,l\ge 1)\nonumber
\end{align}
where $a$ is the t'Hoft coupling $a=g^2N_c/(4\pi^2)$ and $R$ contains all the dependence on the polarization vectors $Y$
\begin{align}
&R =  \frac{y_{12}^2 y_{23}^2 y_{34}^2 y_{41}^2}{x_{12}^2 x_{23}^2 x_{34}^2 x_{41}^2} \bigl( x_{13}^2 x_{24}^2 - x_{12}^2 x_{34}^2 - x_{14}^2 x_{23}^2\bigr) \notag \\
&+ \frac{y_{12}^2 y_{24}^2 y_{43}^2 y_{31}^2}{x_{12}^2 x_{24}^2 x_{43}^2 x_{31}^2} \bigl( x_{14}^2 x_{23}^2 - x_{12}^2 x_{34}^2 - x_{13}^2 x_{24
}^2\bigr) + \frac{y_{13}^2 y_{32}^2 y_{24}^2 y_{41}^2}{x_{13}^2 x_{32}^2 x_{24}^2 x_{41}^2} \bigl( x_{12}^2 x_{34}^2 - x_{13}^2 x_{24}^2 - x_{14}^2 x_{23
}^2\bigr) \notag \\
&+ \frac{y_{12}^4 y_{34}^4}{x_{12}^2 x_{34}^2} + \frac{y_{13}^4 y_{24}^4}{x_{13}^2 x_{24}^2} + \frac{y_{14}^4 y_{23}^4}{x_{14}^2 x_{23}^2}.
\end{align}
The function $f^{(l)}(x_1,\dots,x_{4+l})$ possesses an hidden permutation symmetry $S_{4+l}$ and this, together with imposing the correct OPE behavior, has led to a complete description of its form up to a high loop order\cite{Eden:2011we}. 
An useful representation for  $f^{(l)}(x_1,\dots,x_{4+l})$ is \cite{Eden:2011we,Eden:2012tu}
\begin{align}
f^{(l)}(x_1,\dots,x_{4+l}) = \frac{P^{(l)}(x_1,\dots,x_{4+l})}{\Pi_{1\le i <j \le 4+l}x_{ij}^2}
\end{align}
where $P^{(l)}(x_1,\dots, x_{4+l})$ is a symmetric polynomial depending only on distances $x_{ij}^2$ and is homogeneous of degree $(l-1)(l+4)/2$ in each point. The planar part of this polynomial is given up to four loops by \cite{Eden:2012tu}
\begin{align}
&P^{(1)}=1,\, \ \ \ \ P^{(2)} = \frac{1}{48}x_{12}^2x_{34}^2x_{56}^2+\text{$S_6$ perm},\,P^{(3)} =\frac{1}{20}(x_{12}^2)^2x_{34}^2x_{45}^2x_{56}^2x_{67}^2x_{73}+\text{$S_7$ perm}\nonumber\\
&P^{(4)}_{ g=0} =\frac{1}{24} x_{12}^2 x_{13}^2 x_{16}^2 x_{23}^2 x_{25}^2 x_{34}^2 x_{45}^2 x_{46}^2 x_{56}^2 x_{78}^6+\frac{1}{8}x_{12}^2 x_{13}^2 x_{16}^2
   x_{24}^2 x_{27}^2 x_{34}^2 x_{38}^2 x_{45}^2 x_{56}^4 x_{78}^4
\nonumber\\
 & \ \ \  \ \ \  -\frac{1}{16} x_{12}^2 x_{15}^2 x_{18}^2 x_{23}^2 x_{26}^2 x_{34}^2
   x_{37}^2 x_{45}^2 x_{48}^2 x_{56}^2 x_{67}^2 x_{78}^2+\text{$S_8$ permutations}\label{eq:FourPointInt4loop}
\end{align}
where the $g=0$ is just to remind that this polynomial concerns just the planar sector of the four point function.  The integrals appearing up to three loops can be expressed in terms ladder integrals and two functions called Easy and Hard integrals\cite{Eden:2012tu,Drummond:2013nda}. We have not tried to count the minimum  number of independent integrals that appear at four loop level because the computer time saved is not considerable since we are only interested in the contribution of the Konishi operator. 
\subsection{Asymptotic expansions}
The integrals appearing in the four point function described above are conformal and consequently they depend on two cross ratios $u$ and $v$
\begin{align}
u=\frac{x_{12}^2x_{34}^2}{x_{13}^2x_{24}^2},\, \ \ \ v=\frac{x_{14}^2x_{23}^2}{x_{13}^2x_{24}^2}.
\end{align}
Most of the integrals at four loops are not known explicitly as a function of these cross ratios, however the method of asymptotic expansions can be used to reduce the computation of these four point integrals to the evaluation of simpler integrals involving just two points. The expansion of the integrals in terms of the cross ratios can then be used to extract the dimension and OPE coefficients of the operators that can couple to the external ones. This method  has been used in the past to compute the OPE coefficient of twist two operators at three loops \cite{Eden:2012rr}.

We will review briefly how the method works on $4$ loop integral that appears in this four point function\footnote{More details on the method can be found in \cite{Eden:2012rr,Smirnov:2012gma}.}
\begin{align}
I&=\int\frac{x_{15}^2 x_{67}^2\,d^4x_5d^4x_6d^4x_7d^4x_8}{x_{16}^2 x_{17}^2 x_{18}^2 x_{25}^2 x_{26}^2 x_{27}^2 x_{28}^2 x_{35}^2 x_{36}^2 x_{45}^2 x_{47}^2 x_{56}^2 x_{57}^2 x_{68}^2 x_{78}^2}\nonumber\\
&=\int\frac{x_{5}^2 x_{67}^2\,d^4x_5d^4x_6d^4x_7d^4x_8}{x_{6}^2 x_{7}^2 x_{8}^2 x_{25}^2 x_{26}^2 x_{27}^2 x_{28}^2 x_{35}^2 x_{36}^2  x_{56}^2 x_{57}^2 x_{68}^2 x_{78}^2}.
\end{align}
where the point $x_4$ was sent to infinity and $x_1$ to $0$ using conformal invariance of the integral.
The cross ratios, in these coordinates, are given by $u=x_{2}^2/x_{3}^2$ and $v=x_{23}^2/x_{3}^2$. The method of asymptotic expansions can be used obtain a series expansion in $u$ and $(1-v)$ to any desired order. Powers of $u$ in a four point function control the twist of an operator (recall that twist is defined by $\Delta-J$) and powers of $(1-v)$ control the spin, $J$, of an operator. We are only interested in extracting the OPE coefficient of the Konishi operator, so we can focus only on the leading term of the expansion. 

The main idea of the method is to divide the range of the integration of each integration variable in two regions, one where it is of the order of $x_2\ll 1$, that is assumed to be small, and other where it is of the order of $x_3$. There are four integration variables  and consequently $2^4=16$ integration regions. The goal of dividing into these regions is that it allows to simplify the integrand. For example in the region where all the integration variables are of the order $x_3$  
\begin{align}
\frac{1}{x_{2j}^2}=\sum_{n=0}^{\infty}\frac{(2x_2\cdot x_j-x_2^2)^n}{(x_{j}^2)^{1+n}},\, \ \ \textrm{for } \ \ j=5,\dots ,\,8 .\label{eq:expansiondenominator}
\end{align} 
Obviously, this equation is only valid when the region of integration satisfies $x_{2}^2\le x_j^2$, however we can extend this region of integration to all space at the expense of introducing an regulator $d=4-2\epsilon$. The integrals  will have poles in  $\epsilon$ as a consequence of extending the integration region. However, the sum of all regions needs to give a finite result in the limit of $\epsilon\rightarrow 0$ since we are dealing with finite integrals. 
At the end of the day each integration region is expressed in terms of integrals of the propagator type
\begin{align}
\int & \frac{d^dx_5d^dx_6d^dx_7d^dx_8}{(x_{5}^2)^{a_1}(x_{25}^2)^{a_2}(x_{56}^2)^{a_3}(x_{57}^2)^{a_4}(x_{58}^2)^{a_5}(x_{6}^2)^{a_6}(x_{26}^2)^{a_7}(x_{67}^2)^{a_8}(x_{68}^2)^{a_9}}\times\\
&\times \frac{1}{(x_{7}^2)^{a_{10}}(x_{27}^2)^{a_{11}}(x_{78}^2)^{a_{12}}(x_{8}^2)^{a_{13}}(x_{28}^2)^{a_{14}}}\,,\nonumber \,\, \ \ \ a_i \in\, \mathbb{Z}\label{eq:propagatortypeintegral}
\end{align}
and fortunately all integrals  that are needed have been computed before \cite{Baikov:2010hf,Eden:2012tu}.

Let us go back to the example where all the integration variables are of the order $x_3$ in the integral $I$ 
\begin{align}
\sum_{n_i=0}^{\infty}\int\frac{\prod_{i=5}^{8}(2x_i\!\cdot\! x_2-x_2^2)^{n_i} x_{67}^2\,d^4x_5\dots d^4x_8}{(x_{5}^2)^{n_5}(x_{6}^2)^{2+n_6}( x_{7}^2)^{2+n_7} (x_{8}^2)^{2+n_8} x_{35}^2 x_{36}^2  x_{56}^2 x_{57}^2 x_{68}^2 x_{78}^2}.
\end{align}
Higher powers of $x_2^2$ encode the contribution of higher twist operators, since we are only interested in twist two, we can neglect the factor $x_2^2$ in $(2x_i\!\cdot\! x_2-x_2^2)$. Furthermore, non-zero values of $n_i$ give higher powers of $x_2\cdot x_1$ and these are only required if one is interested in operators with higher spin. So we can focus on the term with $n_i=0$. 

There are also other regions that contribute to this integral, one of them is characterized by $x_5,\,x_6\sim x_2$ and $x_7,\,x_8\sim x_3$. The integrand also simplifies in this case after doing the following changes
\begin{align}
\frac{1}{x_{2j}^2}=\sum_{n=0}^{\infty}\frac{(2x_2\cdot x_j-x_2^2)^n}{(x_{j}^2)^{1+n}}, \ \ \ \frac{1}{x_{3i}^2}=\sum_{n=0}^{\infty}\frac{(2x_3\cdot x_i-x_i^2)^n}{(x_{3}^2)^{1+n}},\ \ \frac{1}{x_{ij}^2}=\sum_{n=0}^{\infty}\frac{(2x_i\cdot x_j-x_i^2)^n}{(x_{j}^2)^{1+n}}
\end{align}
where $i=5,\,6$ and $j=7,\,8$. So, the integrand can be written in this region as
\begin{align}
\!\!\!\!\sum_{n_i,n_{ij}=0}^{\infty}\!\!\frac{x_{5}^2\prod_{j=7}^{8}(2x_2\!\cdot \!x_j-x_2^2)^{n_j}\prod_{j=5}^{6}(2x_3\!\cdot \!x_i-x_i^2)^{n_i}(2x_5\!\cdot \!x_7-x_5^2)^{n_{57}}(2x_6\!\cdot \!x_8-x_8^2)^{n_{68}} x_{67}^2\,}{x_{6}^2x_{56}^2 x_{25}^2 x_{26}^2  x_{78}^2 (x_{3}^2 x_{3}^2)^{2+n_5+n_6}(x_{7}^2)^{3+n_7+n_{57}} (x_{8}^2)^{3+n_8+n_{68}} }.
\end{align}
Notice that the integrals in this region can be viewed as the product of two loop propagators with numerators. In fact this is a feature of the method, an $l$-loop four point conformal integral can be written in terms of $(l-k)$-loop propagator type integral with $k=0,\,\dots, l-1$. 

In this decomposition in terms of simpler integrals it is often the case where one has to deal with integral with open indices or in other words, integral where the numerator is contracted with an external vector. An example of this is 
\begin{align}
\int \frac{x_5^2(x_6^2-2x_6\!\cdot\! x_7+x_7^2)d^dx_5d^dx_6}{x_{6}^2x_{56}^2 x_{25}^2 x_{26}^2}.
\end{align}
These  can be expressed in terms of integrals of the form (\ref{eq:propagatortypeintegral}). The procedure is simple and it is explained in section $3$ of \cite{Eden:2012rr}. 

After all the integrals appearing in the asymptotic expansions are expressed in terms of integrals of the form (\ref{eq:propagatortypeintegral}) one just uses a program such as LiteRed \cite{Lee:2012cn} or FIRE \cite{Smirnov:2013dia} to reduce them to master integrals.  The number of master integrals of the propagator type depends on the loop order, at one, two, three and four loops the number of master integrals is $1$, $5$, $9$ and $24$ respectively.  At this point any conformal integral is given by a combination of master integrals, whose values have been determined in an $\epsilon$ expansion
\begin{align}
j_{k} = \sum_{i=-4}^{\infty}\epsilon^i c_{i,k} \,j_{k,i},\, \ \ k=1,\, \dots, 24. \label{eq:MasterIntegralsExpansion}
\end{align}
up transcendentality seven \cite{Baikov:2010hf,Eden:2012tu}\footnote{Notice that transcendentality is enough for our purposes. In fact, all four loops integrals in momenta space are known up to transcendality $12$ but our integrals are in position space. For planar integrals there is no difference between momenta and position space since they are related by a simple change of variables\cite{Smirnov:2012gma}. However, for non planar there is no simple relation and they have been worked out in  \cite{Eden:2012tu}. }.

\subsection{Konishi from OPE limit}
The OPE decomposition of a four point function can be done in every conformal field theory. In the present case we are interested on the contribution of the Konishi operator $\mathcal{K}(x)$ in the OPE of two $\mathcal{O}(x,y)$ operators
\begin{align}
\mathcal{O}(x_1,y_1)\mathcal{O}(x_2,y_2) = c_{\mathcal{I}}\frac{y_{12}^2}{(x_{12}^2)^2}\mathcal{I}+ c_{\mathcal{K}}\frac{y_{12}^2}{(x_{12}^2)^{1-\gamma_{\mathcal{K}}/2}}\mathcal{K}(x_2)+c_{\mathcal{O}}\frac{y_{12}}{x_{12}^2}Y_1^{I}Y_2^{J}\mathcal{O}_{20}^{IJ}(x_2)+\dots
\end{align}.
The operators that will flow in the four point function depend on the polarizations vectors of the external operators. To understand what operators are exchanged remember that  the tensor product of two $20$'s decomposes in six irreducible representations. It is more convenient to do the OPE in the channel $20$ since there is just one operator flowing with twist two per spin \cite{Dolan:2001tt}. 
Since we are just interested in the of the Konishi we can focus on the leading term in $1-v$ in the light-cone limit
\begin{align}
\sum_{l\ge 1}a^{l}\frac{x_{12}^2x_{13}^2x_{14}^2x_{23}^2x_{24}^2x_{34}^2}{l!(-4\pi^2)^l}\int d^4x_5\dots d^4x_{4+l}f^{(l)}(x_1,\dots,x_{4+l}) \rightarrow &\\
\rightarrow\frac{1}{6x_{13}^{4}}(c_{\mathcal{K}}^2(a)u^{\frac{\gamma_{\mathcal{K}}}{2}}-1)(1+Ou+O(1-v))&\label{eq:OPEOPEEquation}.
\end{align}
The leading term of the integral is given by
\begin{align}
\!\!\!&x_{13}^4\int d^4x_i f^{(4)}(x_i) =4 \left(148 \zeta _3^2+\left(1312-60 \zeta _4\right) \zeta _3+5020 \zeta _5-1250 \zeta _6+8305 \zeta _7+9952\right)\nonumber\\
&-64 \left(78 \zeta _3+55 \left(3 \zeta _5+8\right)\right) \ln u+576 \left(3 \zeta _3+14\right) \ln ^2u-1152 \ln ^3u+72 \ln ^4u+O(u)+O(1-v).\nonumber
\end{align}
The OPE coefficient  can be extracted by comparing the equation above with (\ref{eq:OPEOPEEquation}) where we have used the lower loop data for the anomalous dimension and OPE coefficients 
\footnote{We have used the following values for the OPE and anomalous dimension of the Konishi operator
\begin{align}
&\gamma_{\mathcal{K}}=12 a-48 a ^2+336 a ^3 +256 a ^4 \left(\frac{9 \zeta_3 }{4}-\frac{45 \zeta_5}{8}-\frac{39}{4}\right),\,c_{\mathcal{K}}=\frac{4}{3}-16 a+a ^2 (224+96 \zeta_3 ) -a ^3 (3072+512 \zeta_3 +1600 \zeta_5)\nonumber.
\end{align}
 }
\begin{align}
&c_{\mathcal{K}}^{(4)}= a^4 \,32 \left(36 \zeta_3^2+164 \zeta_3+490 \zeta_5+735 \zeta_7+1244\right).\label{eq:OPECoefficient}
\end{align}

\section{Conclusions}
We have computed, in this short note, the OPE coefficient between two operators $\mathcal{O}_{20'}$ and the Konishi. One of the main motivations to obtain this result was that it will allow to check the integrability computation. The result at four loops is particularly important since in the hexagon approach there is a new effect that will only kick in at this order \cite{Basso:2015eqa}. Recall that it is also at this loop order that the wrapping effects in the spectrum start to contribute. Reproducing the results of this note with the hexagon approach is an important non-trivial check. 

We have focused on the OPE coefficient of the Konishi operator but it is also possible to obtain the OPE coefficients of twist two operators with higher spin. The main hurdle is to decompose a given integral in terms of master integrals. The packages LiteRed and FIRE can do this decomposition but it will demand more computer time. 

There are two more interesting directions, one is the evaluation of the non-planar corrections at four loops and the other is to repeat this procedure for the OPE coefficient but at five loops.

\section*{Acknowledgements} 
We are grateful to Pedro Vieira and Benjamin Basso for encouraging to do this project and for discussions. We would like to thank Gregory Korchemsky  for discussions. V.G.
would also like to thank FAPESP grant 2015/14796-7, CERN/FIS-NUC/0045/2015 and CGI in Florence where this work has been completed. 

\appendix
\section{Master integrals}
Every conformal integral can be expressed as a linear combination of master integrals of the propagator type. These were computed in the literature in momenta space at four loop level \cite{Baikov:2010hf}. A duality between planar integrals in momenta space and position space can be used to determine the coefficients $c_{i,k}$ of (\ref{eq:MasterIntegralsExpansion}) in the planar sector. There are  $4$ more master integrals that could contribute to this four point function
\begin{align}
j_{21}&=\frac{n^4_0(\epsilon)}{\pi^{2d}}\int\frac{d^dx_5d^dx_6d^dx_7d^dx_8}{x_{25}^2x_{56}^2x_{57}^2x_{6}^2x_{68}^2x_{7}^2x_{78}^2x_{28}^2}, \ \ j_{22}=\frac{n^4_0(\epsilon)}{\pi^{2d}}\int\frac{d^dx_5d^dx_6d^dx_7d^dx_8}{x_{25}^2x_{56}^2x_{57}^2x_{6}^2x_{68}^2x_{7}^2x_{78}^2(x_{28}^2)^2}\label{eq:nonplanarmaster}\\
j_{23}&=\frac{n^4_0(\epsilon)}{\pi^{2d}}\int\frac{d^dx_5d^dx_6d^dx_7d^dx_8}{x_{25}^2x_{56}^2x_{57}^2x_{58}^2x_{6}^2x_{67}^2x_{68}^2x_{7}^2x_{27}^2x_8^2x_{28}^2}, \ \ \ \ j_{24}=\frac{n^4_0(\epsilon)}{\pi^{2d}}\int\frac{d^dx_5d^dx_6d^dx_7d^dx_8}{x_5^2x_{25}^2x_{57}^2x_{58}^2x_{6}^2x_{26}^2x_{67}^2x_{68}^2x_{7}^2x_{27}^2x_8^2x_{28}^2}\nonumber
\end{align}
where $n_0(\epsilon)=e^{-\gamma_{e}\epsilon}\Gamma(2-2\epsilon)/(\Gamma(1+\epsilon)\Gamma^2(1-\epsilon))$ converts the integrals to the $G$-scheme \cite{Baikov:2010hf}. The first two integrals were computed in \cite{Eden:2012fe} up to order $\epsilon$ and the other two do not contribute to contribute to the OPE coefficient of the Konishi operator(this should also be true for contribution of the OPE coefficient of other twist two operators with higher spin). 

The method to compute the first two integrals in (\ref{eq:nonplanarmaster}) was presented in the appendix of \cite{Eden:2012fe}. The method is nice and we will review it here. We are interested in evaluating the integrals $j_{21}$ and $j_{22}$ in a Laurent expansion in $2\epsilon=d-4$
\begin{align}
j_{21} = \frac{j_{21,-1}}{\epsilon}+ j_{21,0}+j_{21,1}\epsilon,\, \ \ j_{22} = \frac{j_{21,-1}}{\epsilon}+ j_{21,0}+j_{21,1}\epsilon. 
\end{align}
Imposing that the each conformal integral that appears in the planar part of the four point function (\ref{eq:FourPointInt4loop}) are finite in the limit $\epsilon\rightarrow 0$ and conformal fixes their values\footnote{In our computations we have used $G$-scheme and for the planar master integrals we have used \cite{Baikov:2010hf}.
%
  }
\begin{align}
j_{21,-1}=5\zeta_5,\,  j_{21,0}=\frac{5 \pi ^6}{378}-13 \zeta_3^2-5 \zeta_5, \  \ j_{22,-1}=-20\zeta_5,\,  j_{22,0}=-8 \zeta_3^2+120 \zeta_5 -\frac{10 \pi ^6}{189}\label{eq:firstvaluesfornonplanar}.
\end{align}
Alternatively one can consider the following finite integrals, in order to compute the integrals $j_{21}$ and $j_{22}$,
\begin{align}
\!\!\!\!I_3(\kappa)= \frac{n_0^4(\epsilon)}{\pi^{2d}}\!\!\int\!\! \frac{d^dx_5d^dx_6d^dx_7d^dx_8}{(x_{25}^2x_{28}^2x_{56}^2x_{57}^2x_6^2x_{7}^2x_{67}^2x_{68}^2x_{78}^2)^{1-\epsilon \kappa}},\,  I_4(\kappa)= \frac{n_0^4(\epsilon)}{\pi^{2d}}\!\!\int\!\! \frac{d^dx_5d^dx_6d^dx_7d^dx_8}{(x_{25}^2x_{26}^2x_{28}^2x_{56}^2x_{57}^2x_6^2x_{7}^2x_{68}^2x_{78}^2)^{1-\epsilon \kappa}}\nonumber.
\end{align}
Both $I_3$ and $I_4$ admit a power series in $\epsilon$
\begin{align}
I_{i}(\kappa) = b_i+\epsilon (c_i+\kappa d_i) +O(\epsilon^2).
\end{align}
It turns out that it is easier to evaluate these integrals than $j_{21}$ and $j_{22}$  for particular values of $\kappa$. Then we use the fact that the power series expansion in $\epsilon$ is linear in $\kappa$ at first order in  $\epsilon$ to obtain $b_i$ and $c_i$. These constants are related to $j_{21,0},\, j_{21,1},\, j_{22,0}$ and $j_{22,1}$ by 
integration by parts\footnote{We have used LiteRed \cite{Lee:2012cn} package to do this reduction. There are terms in following equation that are different from \cite{Eden:2012fe}, we think that this might be related to a different conventions. However, we were able to verify the four loop anomalous dimension which also gives us some confidence of the correctness our result.}
\begin{align}
&b_3+c_3\,\epsilon =\left(\frac{830 \zeta_5}{3}-\frac{2 j_{21,0}}{3}-\frac{7 j_{22,0}}{3}+\frac{26 \zeta_3^2}{3}-\frac{65 \pi ^6}{567}\right)\label{eq:IBPNonplanar}+\\
&\epsilon  \left(\frac{14 j_{21,0}}{3}+\frac{14 j_{22,0}}{3}-\frac{2 j_{21,1}}{3}-\frac{7 j_{22,1}}{3}-\frac{208 \zeta_3^2}{3}+\frac{13 \pi ^4 \zeta_3}{45}-\frac{3220 \zeta_5}{3}-\frac{4667 \zeta_7 }{6}+\frac{520 \pi ^6}{567}\right)+\nonumber\\
&b_4+c_4\,\epsilon =\left(235 \zeta_5-j_{21,0}-2 j_{22,0}+7 \zeta_3^2-\frac{5 \pi ^6}{54}\right)\\
&+\epsilon  \left(2 j_{21,0}-j_{21,1}-6 j_{22,0}-2 j_{22,1}-21 \zeta_3^2+\frac{7 \pi ^4 \zeta_3}{30}+285 \zeta_5-\frac{4193 \zeta_7 }{4}+\frac{5 \pi ^6}{18}\right).\nonumber
\end{align}
We will use the values $\kappa =1 $ and $\kappa =\frac{1}{2}$. These have been computed  in \cite{Eden:2012fe}
\begin{align}
I_3\left(\frac{1}{2}\right)&=I_4\left(\frac{1}{2}\right)= n_0^4(\epsilon)\big[\left(144 \zeta_3^2+108 \zeta_4 \zeta_3\right) \epsilon +36 \zeta_3^2\big]+O(\epsilon^2)\\
I_3(1)&=n_0^4(\epsilon)\big[\left(288 \zeta_3^2+108 \zeta_4 \zeta_3-378 \zeta_7\right) \epsilon +36 \zeta_3^2\big]+O(\epsilon^2),\, \ \\
I_4(1)&=n_0^4(\epsilon)\big[\left(108 \zeta_3^2+108 \zeta_4 \zeta_3+\frac{189 \zeta_7}{2}\right) \epsilon +36 \zeta_3^2\big]+O(\epsilon^2)
\end{align}
The constants $b_i$ and $c_i$ are obtained using
\begin{align}
&2I_i\left(\frac{1}{2}\right)-I_i\left(1\right)=b_i+\epsilon c_i\\
&b_3=36\zeta_3^2,\, \ \ c_3=\frac{6}{5} \left(315 \zeta_7 -240 \zeta_3^2+\pi ^4 \zeta_3 \right)\\
&b_4=36\zeta_3^2,\, \ \ d_4=\frac{3}{10} \left(4 \pi ^4 \zeta_3 -360 \zeta_3^2-315 \zeta_7\right).
\end{align}
Plugging these values in (\ref{eq:IBPNonplanar}) we obtain again (\ref{eq:firstvaluesfornonplanar}) and also the values for $j_{22,1}$ and $j_{22,1}$
\begin{align}
&j_{21,1}=13 \zeta_3^2-\frac{13 \pi ^4 \zeta_3}{30}+35 \zeta_5+\frac{345 \zeta_7}{4}-\frac{5 \pi ^6}{378}\\
&j_{22,1}=48 \zeta_3^2-\frac{4 \pi ^4 \zeta_3}{15}-240 \zeta_5-520 \zeta_7+\frac{20 \pi ^6}{63}\label{eq:lastvaluesformasterintegrals}.
\end{align}

\end{document}